\documentclass[12pt]{article}
\usepackage{graphics}
\usepackage{amssymb,epsfig,amsmath,euscript,array}
\usepackage{axodraw4j}
\usepackage{subfigure}
\usepackage{color}
\usepackage{cite}

\makeatletter
\@addtoreset{equation}{section}
\makeatother

\definecolor{holger}{rgb}{0,0.5,0.7}
\definecolor{edit}{rgb}{1,0,0}

\definecolor{durbeer}{rgb}{1,0,0}

\definecolor{durbeer2}{rgb}{0.8,0,0.5}

\newcommand{\Eqref}[1]{Eq.~\eqref{#1}}

\newcommand{\charge}{\varepsilon}
\newcommand{\Bmag}{{\mathbf{B}}}

\newcommand{\pbot}{\mathbf{p}_\bot}
\newcommand{\alphat}{\tilde{\alpha}}

\newcounter{multieqs}

\def\bd{\begin{document}}
\def\ed{\end{document}}
\def\nn{\nonumber}
\def\bea{\begin{eqnarray}}
\def\eea{\end{eqnarray}}
\let\bm=\bibitem
\let\la=\label

\begin{document}

\hfill{IPPP/09/21; DCPT/09/42}\\[-0.9cm]

\vspace{20pt}

\begin{center}

{\huge \bf  Tunneling of the 3rd kind} \\[1.5ex]

\vspace{30pt}

{\bf Holger Gies$^1$ and Joerg Jaeckel$^{2}$}

{\small \em
{}$^1$Theoretisch-Physikalisches Institut, Friedrich-Schiller-Universit¨at Jena,\\
Max-Wien-Platz 1, D-07743 Jena, Germany\\[1ex]
{}$^2${Institute for Particle Physics Phenomenology, Durham
University,\\ Durham DH1 3LE, United Kingdom}}

\vspace{10pt}

{\sffamily \tt
holger.gies@uni-jena.de\\
joerg.jaeckel@durham.ac.uk}

\vspace{30pt}
\end{center}

\begin{abstract}
  We study a new kind of tunneling of particles through a barrier particular
  to quantum field theory.  Here, the particles traverse the barrier by
  splitting into a virtual pair of particles of a different species which
  interacts only very weakly with the barrier and can therefore pass through
  it. Behind the barrier, the pair recombines into a particle of the original
  species.  As an example, we discuss the case where photons split into a pair
  of minicharged particles.  This process could {be observed} in experiments of
  the light-shining-through-a-wall type and may be used to search for
  minicharged particles in laboratory experiments.
\end{abstract}

\setcounter{page}{0}
\thispagestyle{empty}
\newpage


\section{Introduction}
Tunneling of particles through a potential barrier is a paradigmatic quantum
mechanical process \cite{bib:tunneling1,bib:tunneling2}. A particle may cross
a potential barrier of finite height and thickness because it can penetrate
into classically forbidden regions with finite probability in agreement with
Heisenberg's uncertainty principle.  Accordingly, the amplitude for the tunneling
process is controlled by the Planck constant $\hbar$, the height $\Delta E$
and width $\Delta x$ of the potential barrier, $\sim\exp(-\sqrt{2m \Delta
  E}\Delta x/\hbar)$.

\begin{figure}
\begin{center}
\scalebox{0.7}[0.7]{
\fcolorbox{white}{white}{
  \begin{picture}(466,194) (47,-111)
    \SetWidth{1.0}
    \SetColor{Black}
    \Photon(48,-14)(160,-14){7.5}{6}
    \GBox(256,-110)(288,82){0.882}
    \Photon(384,-14)(512,-14){7.5}{6}
    \Line[dash,dashsize=10,arrow,arrowpos=0.5,arrowlength=5,arrowwidth=2,arrowinset=0.2](160,-14)(384,-14)
    \Vertex(160,-14){5.657}
    \Vertex(160,-14){7.071}
    \Vertex(384,-14){7.071}
  \end{picture}
}
}
\end{center}
\caption{Diagram depicting a classical process for a penetration of the
  barrier via conversion into a real particle that interacts only very weakly
  with the barrier.} \label{classical}
\end{figure}
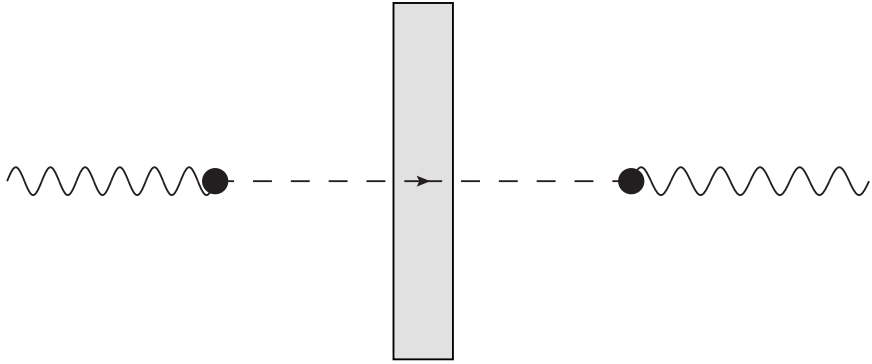

Field theory with different particle species allows for a different
way to penetrate, or more precisely circumnavigate, a barrier.  The
particle can transmute or oscillate into a different species that
does not (or only very weakly) interact with the barrier. Behind the
barrier, the particle then reconverts into the original species.
This process is depicted in Fig.~\ref{classical} and {forms the
basis~\cite{Sikivie:1983ip} of so-called
light-shining-through-a-wall experiments~\cite{Ehret:2007cm} that
can be used to search for axions~\cite{Sikivie:1983ip} and other
light particles~\cite{Okun:1982xi,Ahlers:2007rd}.}  Since the
intermediate particles are real and do not interact with the
barrier, height and width of the original barrier do not matter.

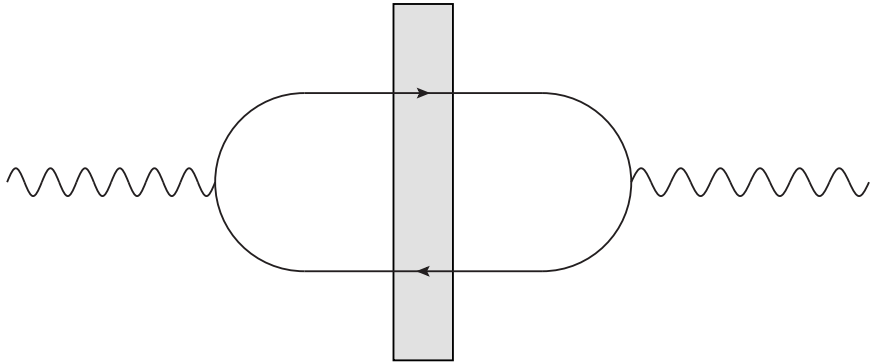
\begin{figure}
\begin{center}
\scalebox{0.7}[0.7]{
\fcolorbox{white}{white}{
  \begin{picture}(466,194) (47,-111)
    \SetWidth{1.0}
    \SetColor{Black}
    \Photon(48,-14)(160,-14){7.5}{6}
    \GBox(256,-110)(288,82){0.882}
    \Arc[clock](208,-14)(48,-90,-270)
    \Line[arrow,arrowpos=0.5,arrowlength=5,arrowwidth=2,arrowinset=0.2](208,34)(336,34)
    \Line[arrow,arrowpos=0.5,arrowlength=5,arrowwidth=2,arrowinset=0.2,flip](208,-62)(336,-62)
    \Arc[clock](336,-14)(48,90,-90)
    \Photon(384,-14)(512,-14){7.5}{6}
  \end{picture}
}
}
\end{center}
\caption{Diagram depicting ``tunneling of the 3rd kind''. The photon splits
  into a virtual pair of particle and antiparticle which traverse the wall and
  recombine into a photon.} \label{tunneling}
\end{figure}

In this note we study a generalization of this process allowed by {\emph{quantum}}
field theory in which the intermediate particle(s) crossing the barrier are
not real but \emph{virtual}. For example, the initial particle could
split into a virtual particle-antiparticle pair which then recombines behind
the barrier as depicted in Fig.~\ref{tunneling}.  Again, the intermediate
particles do not interact with the barrier so the height of the barrier does
not matter. The width of the barrier, however, matters because the
virtuality of the intermediate particles typically goes along with a
characteristic length scale.

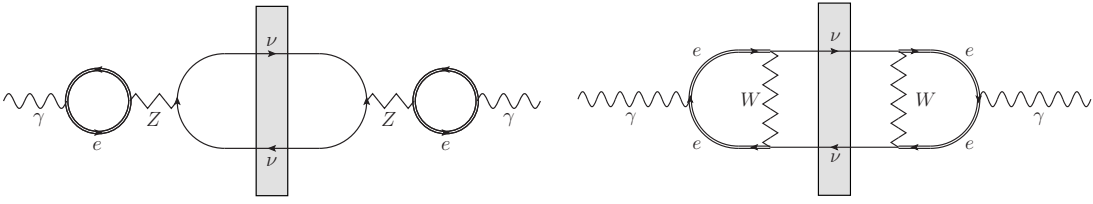
\begin{figure}
\begin{center}
\subfigure{\scalebox{0.45}[0.45]{
\fcolorbox{white}{white}{
  \begin{picture}(451,160) (0,-91)
    \SetWidth{1.0}
    \SetColor{Black}
    \GBox(211.587,-90.916)(238.035,67.774){0.882}
    \Photon(0,-11.571)(52.897,-11.571){6.199}{3}
    \Arc[arrow,arrowpos=0.75,arrowlength=5,arrowwidth=2,arrowinset=0.2,double,sep=2](79.345,-11.571)(26.448,0,360)

    \Line[arrow,arrowpos=0.5,arrowlength=5,arrowwidth=2,arrowinset=0.2](185.139,28.101)(264.484,28.101)
    \Line[arrow,arrowpos=0.5,arrowlength=5,arrowwidth=2,arrowinset=0.2,flip](185.139,-51.244)(264.484,-51.244)
    \Arc[arrow,arrowpos=0.75,arrowlength=5,arrowwidth=2,arrowinset=0.2,double,sep=2](370.277,-11.571)(26.448,0,360)
    \Photon(396.726,-11.571)(449.623,-11.571){6.199}{3}
    \ZigZag(105.794,-11.571)(145.466,-11.571){6.199}{2}
    \ZigZag(304.156,-11.571)(343.829,-11.571){6.199}{2}
    \Arc[arrow,arrowpos=0.5,arrowlength=5,arrowwidth=2,arrowinset=0.2,clock](185.139,-11.571)(39.673,-90,-270)
    \Arc[arrow,arrowpos=0.5,arrowlength=5,arrowwidth=2,arrowinset=0.2](264.484,-11.571)(39.673,-90,90)
    \Text(25,-33)[lb]{\Large{\Black{$\gamma$}}}
    \Text(75,-53)[lb]{\Large{\Black{$e$}}}
    \Text(120,-33)[lb]{\Large{\Black{$Z$}}}
    \Text(221,35)[lb]{\Large{\Black{$\nu$}}}
    \Text(221,-66)[lb]{\Large{\Black{$\nu$}}}
    \Text(317.381,-33)[lb]{\Large{\Black{$Z$}}}
    \Text(368,-53)[lb]{\Large{\Black{$e$}}}
    \Text(420,-33)[lb]{\Large{\Black{$\gamma$}}}
    \Arc[arrow,arrowpos=0.25,arrowlength=5,arrowwidth=2,arrowinset=0.2,double,sep=2](79.345,-11.571)(26.448,0,360)
    \Arc[arrow,arrowpos=0.25,arrowlength=5,arrowwidth=2,arrowinset=0.2,double,sep=2](370.277,-11.571)(26.448,0,360)
  \end{picture}
}
}}
\subfigure{\scalebox{0.43}[0.43]{
\fcolorbox{white}{white}{
  \begin{picture}(451,170) (13,-97)
    \SetWidth{1.0}
    \SetColor{Black}
    \Photon(365.235,-12.292)(463.567,-12.292){6.585}{6}
    \GBox(224.76,-96.576)(252.855,71.993){0.882}
    \Arc[arrow,arrowpos=0.5,arrowlength=5,arrowwidth=2,arrowinset=0.2,clock,double,sep=2](323.092,-12.292)(42.142,90,-90)
    \Arc[arrow,arrowpos=0.5,arrowlength=5,arrowwidth=2,arrowinset=0.2,clock,double,sep=2](154.522,-12.292)(42.142,-90,-270)
    \Photon(14.047,-12.292)(112.38,-12.292){6.585}{6}
    \Line[arrow,arrowpos=0.5,arrowlength=5,arrowwidth=2,arrowinset=0.2,double,sep=2](154.522,29.851)(182.617,29.851)
    \Line[arrow,arrowpos=0.5,arrowlength=5,arrowwidth=2,arrowinset=0.2](182.617,29.851)(294.997,29.851)
    \Line[arrow,arrowpos=0.5,arrowlength=5,arrowwidth=2,arrowinset=0.2,double,sep=2](294.997,29.851)(323.092,29.851)
    \Line[arrow,arrowpos=0.5,arrowlength=5,arrowwidth=2,arrowinset=0.2,double,sep=2](323.092,-54.434)(294.997,-54.434)
    \Line[arrow,arrowpos=0.5,arrowlength=5,arrowwidth=2,arrowinset=0.2](294.997,-54.434)(182.617,-54.434)
    \Line[arrow,arrowpos=0.5,arrowlength=5,arrowwidth=2,arrowinset=0.2,double,sep=2](182.617,-54.434)(154.522,-54.434)
    \ZigZag(182.617,29.851)(182.617,-54.434){6.585}{5}
    \ZigZag(294.997,29.851)(294.997,-54.434){6.585}{5}
    \Text(56,-35)[lb]{\Large{\Black{$\gamma$}}}
    \Text(415,-35)[lb]{\Large{\Black{$\gamma$}}}
    \Text(157,-18)[lb]{\Large{\Black{$W$}}}
    \Text(311,-18)[lb]{\Large{\Black{$W$}}}
    \Text(236,-68)[lb]{\Large{\Black{$\nu$}}}
    \Text(236,39)[lb]{\Large{\Black{$\nu$}}}
    \Text(115,-56)[lb]{\Large{\Black{$e$}}}
    \Text(115,26)[lb]{\Large{\Black{$e$}}}
    \Text(355,-56)[lb]{\Large{\Black{$e$}}}
    \Text(355,26)[lb]{\Large{\Black{$e$}}}
  \end{picture}
}
}}
\end{center}
\caption{In a magnetic background field, tunneling of the 3rd kind {can
    occur} via the effective field-induced interaction between photons
  and neutrinos~\cite{Gies:2000wc} (we have depicted the electron propagator
  in the background field by a double line).  In this case, the neutrinos
  {play} the role of the particles which do not interact with the
  barrier. Using dimensional arguments one can estimate this effect to be {of
    order} \mbox{$\sim(\alpha^2 G_{F}^{2} B^{2} \omega^4 /m_{e}^{4}) ^2
    F(d,m_{\nu},\omega)\lesssim10^{-130}F(d,m_{\nu},\omega)$} where the
  function $F$ parameterizes the dependence on the wall thickness $d$ and the
  neutrino mass $m_{\nu}$ and the right hand side holds for $\omega\sim1\,{\rm
    eV}$ and magnetic fields in the $1\,{\rm T}$ range.} \label{neutrino}
\end{figure}

As a concrete example of this ``tunneling of the 3rd kind'', we study the case
of a photon splitting into a pair of particles with a tiny electric charge,
so-called minicharged particles, in Sects.~\ref{setup}, \ref{results}. The
barrier can be thought of as a mirror or, alternatively, an opaque wall. If
the charges of the intermediate particles are small enough, these minicharged
particles have a tiny cross section with the atoms in the mirror and
consequently simply pass through the wall. By contrast, the analogous process
with electrons does not work, because electrons, {too,} would interact with the
mirror/wall and would be stopped.  Moreover, as we shall see below, for photon
frequencies below the mass of the created virtual particles the process is
exponentially suppressed.  {Another possibility within the standard model
  would be the conversion of the photon into a neutrino--antineutrino pair,
  {for instance, as can be stimulated by a magnetic field, see}
  Fig.~\ref{neutrino}. But since neutrinos couple to photons only very
  indirectly, this process is highly suppressed.}  Therefore, the
standard model background for the corresponding
light-shining-through-a-wall signal is very small and tunneling of
the 3rd kind can serve as a tool to search for physics beyond the
standard model in the form of light minicharged particles. The
latter arise naturally {and consistently\footnote{{In theories with
kinetic mixing~\cite{Holdom:1985ag} minicharged particles are
indeed consistent with the existence of magnetic
monopoles~\cite{Brummer:2009cs}.}}} in many extensions of the
standard model based on field and string
{theory~\cite{Holdom:1985ag,Dienes:1996zr}.}

A particularly interesting type of experiment to search for a
light-shining-through-a-wall effect caused by tunneling of the 3rd kind is the
``superconducting box'' experiment proposed in~\cite{Jaeckel:2008sz} in which
one searches for magnetic fields leaking into a volume shielded by a
superconductor. We will look at this option in Sect.~\ref{box} and estimate
the sensitivity for such an experiment.

\section{Setting}\label{setup}

Let us identify the ingredients for constructing the transition
amplitude and probability for a particle to go through the wall via tunneling of
the 3rd kind as depicted in Fig.~\ref{tunneling}. To be explicit, we
concentrate on the case of a photon fluctuating into two minicharged
particles. The generalization to other types of particles is, however,
straightforward.

Including quantum effects, the effective Lagrangian for a propagating photon
is given by:
\begin{equation}
\mathcal{L}[A]= -\frac{1}{4} F_{\mu\nu}(x) F^{\mu\nu}(x) -
\frac{1}{2}\int_{x'}  A_\mu(x) \Pi^{\mu\nu}(x,x') A_\nu(x'),\label{eq:calL}
\end{equation}
where $\Pi^{\mu\nu}(x,x')$ denotes the two-point correlator, i.e., the vacuum
polarization tensor.

The matter in the wall modifies the `classical' first part of the Lagrangian
\eqref{eq:calL} by boundary conditions such that the photon classically cannot
cross the wall.  On the one-loop level, the vacuum polarization tensor (or,
more precisely, the contribution generated by the minicharged particles)
arises from the loop of minicharged particles shown in
Fig.~\ref{tunneling}. Since minicharged particles interact only very weakly
with the matter of the wall, this part of the vacuum polarization tensor
remains essentially unaffected by the presence of the wall and allows for a
non-vanishing transition amplitude for photons through the wall. In the
following, we will calculate the transition amplitude caused by the
polarization tensor in the presence of boundary conditions arising from the
wall.

If translational invariance holds for the fluctuations
(not necessarily for the photon field) the resulting polarization tensor
satisfies $\Pi^{\mu\nu}(x,x')=\Pi^{\mu\nu}(x-x')$. Together with the Ward
identity, this implies that the polarization tensor can be written in terms of
a single scalar function in momentum space,
\begin{equation}
\Pi_{\mu\nu}(p)=P_{\mathrm{T},\mu\nu}(p)\, \Pi(p), \quad
P_{\mathrm{T},\mu\nu}(p)=g_{\mu\nu}-\frac{p_\mu p_\nu}{p^2}. \label{eq:Pi}
\end{equation}
The metric is given by $g=(-,+,+,+)$, such that $p^2=-\omega^2 + \mathbf{p}^2$.
The equation of motion resulting from \Eqref{eq:calL} for transversal modes
$A_{\mathrm{T},\mu}= P_{\mathrm{T},\mu\nu}A^{\nu}$ reads in momentum space
\begin{equation}
\big(p^2 + \Pi(p) \big) A_{\mathrm{T}}(p)=0.
\end{equation}
Here and in the following, we drop Lorentz indices, since our considerations
are independent of the polarization of the transversal mode.  In this work, we
consider a set-up where translational invariance is broken for the photon
field along the $z$ axis. Hence, it is useful to introduce the partial Fourier
transforms ($p^2=-\omega^2 +\pbot^2 +p_z^2$),
\begin{eqnarray}
A(z,\pbot,\omega)&=& \int \frac{d p_z}{2\pi} \, e^{i z p_z}\,
A(p), \label{eq:FTA}\\
\Pi(z-z',\pbot,\omega)&=& \int \frac{d p_z}{2\pi} \, e^{i (z-z') p_z}\,
\Pi(p), \label{eq:FTPi}
\end{eqnarray}
in terms of which the equations of motion read
\begin{eqnarray}
0&=& (-\omega^2 +\pbot^2- \partial_z^2) A_{\mathrm{T}}(z,\pbot,\omega) + \int dz'\,
\Pi(z-z',\pbot,\omega)\, A_{\mathrm{T}}(z',\pbot,\omega)\nonumber\\
&\equiv& (-\omega^2+\pbot^2- \partial_z^2) A_{\mathrm{T}}(z,\pbot,\omega)
+j(z,\pbot,\omega). \label{eq:j}
\end{eqnarray}
In the last step, we have introduced the fluctuation-induced current
$j=\int\Pi A_{\mathrm{T}}$.

In the present work, we break translational invariance for the photon by a
wall of thickness $d$, infinitely extended into the $x,y$ plane. The left side
of the wall is put at $z=0$, the right side extends to $z=d$. The wall imposes
boundary conditions on the photon field. For instance, if the wall is
perfectly conducting, a transverse photon propagating along the $z$ axis
normal to the wall ($\pbot=0$) has to satisfy Dirichlet boundary conditions at
the wall's surface, corresponding to vanishing transverse electric components
on the conductor. For a wave packet $a(\omega)$, the free equation of motion
for the left half-space $z\leq0$ is solved by
\begin{equation}
A_{\mathrm{T}}(z,\omega)=a(\omega)\, \sin(\omega z), \quad z<0. \label{eq:Ainit}
\end{equation}
In absence of any external photon field to the right of the wall $z\geq d$,
the induced current in this region of space is
\begin{equation}
j(z>0,\omega) = \int_{-\infty}^0 dz'\, \Pi(z-z',\omega)\, a(\omega)\, \sin
(\omega z'). \label{eq:jright}
\end{equation}
We observe that the quantum nonlocalities, or loosely speaking, the spatial
extent of the fluctuations described by the two-point correlator {$\Pi(x,x')$}
give rise to a nonvanishing source on the right hand side of the wall.

The solution to the free Green's function equation,
\begin{equation}
(-\omega^2 -\partial_z^2) G(z-z') = \delta (z-z'), \label{eq:G}
\end{equation}
for $z>z'$ reads
\begin{equation}
\label{greens}
G(z-z') = \frac{i}{2 \omega} \, e^{i \omega (z-z')},
\end{equation}
such that the induced outgoing wave to the right of the wall is given by
\begin{equation}
A_{\mathrm{ind}}(z\gg d, \omega)= i \int_d^\infty dz'\, \frac{e^{i\omega (z-z')}}{2\omega} \,
j(z',\omega). \label{eq:Aind}
\end{equation}
In the present case, the polarization of $A_{\mathrm{ind}}$ is identical to that
of the incident photon.  In \Eqref{eq:Aind}, we have confined ourselves to the
outgoing right-moving far field at $z\gg d$. Near the wall, the Green's
function in the presence of the boundary at $z=d$ has to be used instead; the
latter in addition contains left-moving components which are of no relevance
in the following.

The transition probability for a photon to cross the wall is given by the
square of the photon amplitude normalized to the initial amplitude.  Using
Eqs.~\eqref{eq:Aind} and \eqref{eq:jright}, we find,
\begin{equation}
\label{transition}
P_{\gamma\to\gamma}=\lim_{z\to\infty} \left|\frac{A_{\rm ind}(z,\omega)}{a(\omega)}\right|^2
=\frac{1}{4\omega^2}\left|\int^{\infty}_{d}dz^{\prime}\int^{0}_{-\infty} dz^{\prime\prime}
  \Pi(z^{\prime}-z^{\prime\prime},\omega)\sin(\omega z^{\prime\prime})\exp(-i\omega z^{\prime})\right|^2.
\end{equation}
Generalizations to different boundary conditions for the photon field at the
barrier are straightforward.

\section{Photon transition amplitude}\label{results}

Let us assume the existence of minicharged particles that couple weakly to
photons but not directly to matter which the wall consists of. For minimally
coupled minicharged Dirac fermions, the well-known results from QED can
immediately be taken over.  We begin with the well-known {representation\footnote{{In the Appendix,
we check our calculation by using the Feynman parameter representation of the polarization tensor.}}} of the
polarization tensor in QED, see, e.g., \cite{Dittrich}, where\footnote{Our
  conventions for $\Pi(p)$ differ from those of \cite{Dittrich} by an
  additional factor of $p^2$.}
\begin{equation}
\Pi(p)=-\frac{\alphat}{3\pi}\, p^4 \int_0^1 dv\, \frac{v^2 (3-v^2)}{1-v^2} \,
\frac{1}{p^2+ \frac{4 m^2}{1-v^2} -i \epsilon}.
\end{equation}
Here, $\alphat$ is the analogue of the QED coupling constant including the minicharge $\varepsilon\ll 1$,
$\alphat=\charge^2 \alpha$, $\alpha\simeq 1/137$.

The $p_z$ integral can be done by using the residue theorem, exhibiting two
poles: one below and one above the real $p_z$ axis. Since we are eventually
interested in the polarization tensor for $z>0$, we close the contour in the
upper $p_z$ half plane and pick up the residue of the corresponding pole. Let
us carefully distinguish between the following two cases: For large
frequencies, $\omega > \pbot^2 + \frac{4 m^2}{1-v^2}$, the pole occurs close
to the real axis at
\begin{equation}
p_z= \sqrt{\omega^2 -\pbot^2 -\frac{4m^2}{1-v^2}} + \frac{1}{2}
\frac{i \epsilon}{\sqrt{\omega -\pbot^2 - \frac{4m^2}{1-v^2}}} +
\mathcal{O}(\epsilon^2)
\end{equation}
For small frequencies, $\omega < \pbot^2 + \frac{4
    m^2}{1-v^2}$, the pole lies on the imaginary axis,
\begin{equation}
p_z =i\sqrt{\pbot^2 + \frac{4m^2}{1-v^2} - \omega}.
\end{equation}
As a result, we obtain for the two cases:
\begin{eqnarray}
\Pi(z,\pbot,\omega) &=& -  i \frac{\alphat}{3\pi}
 \int_0^1 dv \frac{v^2 (3-v^2)}{1-v^2}\, \left( \frac{4
    m^2}{1-v^2} \right)^2  \nonumber\\
&&\cdot \left\{ \begin{array}{ll}
\frac{1}{2\sqrt{\omega^2 -\pbot^2 - \frac{4m^2}{1-v^2}}} e^{i z
  \sqrt{\omega^2 -\pbot^2 - \frac{4m^2}{1-v^2}}} & \quad \text{for}\,\,
\omega^2> \pbot^2 + \frac{4m^2}{1-v^2} \\
\frac{1}{2i \sqrt{\pbot^2 + \frac{4m^2}{1-v^2} -\omega^2 }} e^{- z
  \sqrt{\pbot^2 + \frac{4m^2}{1-v^2} -\omega^2 }} & \quad \text{for}\,\,
\omega^2< \pbot^2 + \frac{4m^2}{1-v^2}
\end{array}\right. .\label{eq:Picase}
\end{eqnarray}
This representation can now be plugged into Eq.~(\ref{transition}). For
simplicity, we confine ourselves to photon propagation parallel to the $z$
axis with $\pbot=0$. For frequencies below threshold, $\omega< 2m$, only the
second case occurs; above threshold, both cases contribute.
Using the substitutions
\begin{equation}
\lambda=\sqrt{1- \frac{4m^2}{\omega^2(1-v^2)}}, \quad \kappa =
\sqrt{\frac{4m^2}{\omega^2(1-v^2)} -1},
\end{equation}
for the first and second case, respectively, the induced outgoing photon field
can be computed from Eqs.~(\ref{eq:jright}) and (\ref{eq:Aind}), yielding the
representation
\begin{equation}
A_{\mathrm{ind}}(\omega)= \frac{i \alphat}{6\pi}\, a(\omega) e^{i\omega (z-d)}
  \Big( f_>(\omega/m, \omega d) + f_<(\omega/m, \omega d) \Big),
\end{equation}
where we have introduced the dimensionless auxiliary functions
\begin{eqnarray}
f_>(\omega/m,\omega d) &=& \int_0^{\mathrm{Re}\sqrt{1- \frac{4m^2}{\omega^2}}}
\frac{d\lambda}{1-\lambda} \frac{
  \sqrt{1-\lambda^2 -\frac{4m^2}{\omega^2}} }{\sqrt{1-\lambda^2}}
\frac{\left( 1-\lambda^2 + \frac{2m^2}{\omega^2}
  \right)}
{1-\lambda^2}
e^{i \omega d\lambda},\label{eq:3}\\
f_<(\omega/m,\omega d) &=& \int_{\mathrm{Re}\sqrt{\frac{4m^2}{\omega^2}-1}}^\infty
\,\frac{d\kappa}{i+\kappa}
\frac{\sqrt{1+\kappa^2 - \frac{4m^2}{\omega^2}}}{\sqrt{1+\kappa^2}}
\frac{\left( 1+\kappa^2 + \frac{2m^2}{\omega^2}\right)}{1+\kappa^2}
e^{-\omega d \kappa}.
\end{eqnarray}
These auxiliary functions can be numerically evaluated to a high precision
with standard routines. Insertion of the result into Eq.~(\ref{transition})
yields the final tunneling probability,
\begin{equation}
P_{\gamma\to\gamma}= \frac{\alphat^2}{36 \pi^2} \big| f_> + f_< \big|^2,
\label{eq:1}
\end{equation}
which will be discussed in various limiting cases in the following.

\subsection{Small-frequency limit $\omega\ll 2 m$}

\begin{figure}[t]
\begin{center}
\includegraphics[angle=0,width=.6\textwidth]{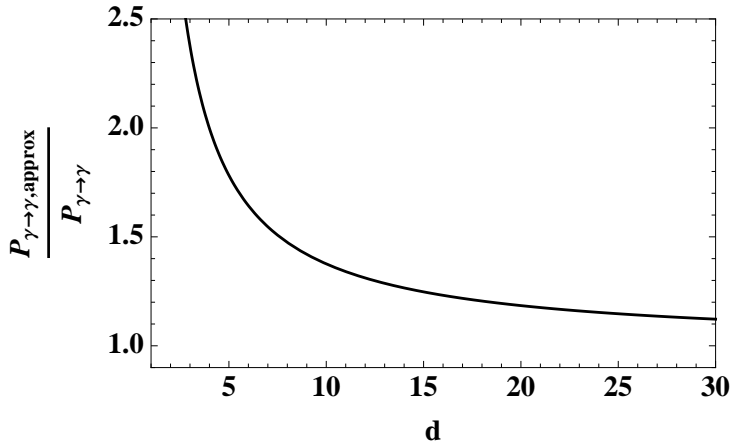}
\end{center}
\vspace{-0cm} \caption{{Ratio of the approximate expression for the transition
    probability $P_{\gamma\to\gamma}(d)$ of a photon through a wall of
    thickness $d$ given in Eq.~\eqref{eq:2} to the exact evaluation of
    Eq.~\eqref{smallf}.  We have used $\omega=0$, $m=1$.}}
\label{compare1}
\end{figure}

For all frequencies $\omega<2m$, we have $f_>=0$ such that only the function
$f_<$ contributes. Rescaling the integration variable $\kappa$ such that the
lower bound of the $f_<$ integral is unity, the limit $\omega\ll 2 m$ reduces to
a simpler representation:
\begin{equation}
\label{smallf}
f_<(\omega\ll m) =\int_1^\infty \frac{d\kappa}{\kappa^4}\,  \sqrt{\kappa^2-1} \,
\left(\kappa^2+\frac{1}{2}\right)\, e^{-2  md \kappa}.
\end{equation}
In the limit $md\gg 1$ corresponding to thick walls compared to the Compton
wavelength of the minicharged particle, the asymptotics of the integral can be
extracted by a saddle-point approximation, yielding
\begin{eqnarray}
f_<(\omega\ll 2 m, md\gg 1) &\simeq& \frac{3\sqrt{\pi}}{8 (md)^{\frac{3}{2}}} \, e^{-2md},
\nonumber\\
P_{\gamma\to\gamma}(md \gg 1, \omega\ll 2m)
&\simeq&\frac{\alphat^2}{256 \pi} \frac{e^{-4md}}{(md)^3}
=\frac{\charge^4\alpha^2}{256 \pi} \frac{e^{-4md}}{(md)^3}. \label{eq:2}
\end{eqnarray}
This results exhibits a typical exponential decrease with an exponent which is
linear in the wall thickness $d$, as is familiar from quantum mechanical
tunneling processes. In Fig.~\ref{compare1}, we compare the approximate
result, Eq.~\eqref{eq:2}, for the transition probability of photons through
the wall to an exact numerical evaluation of the integrals. Already for modest
values of the wall thickness, {$m\,d\gtrsim 20$,} we find reasonable
quantitative agreement on the level of $25\%$.

The analogy to quantum mechanical tunneling becomes even more transparent in
the worldline approach to quantum field theory \cite{Schubert:2001he}. Here,
quantum field theoretic propagators are represented by quantum mechanical path
integrals in a fictious time. These paths can be thought of as
Lorentz-invariant spacetime trajectories of the quantum fluctuations. In the
above limit of small photon frequency and large wall thickness, these path
integrals can be approximated semiclassically, resulting in
minicharged-particle propagators of the form $G(x-y)\simeq
\sqrt{\pi/[2m(x-y)]} \exp [-m(x-y)]$. As the probability amplitude for our
process involves two {propagators, the probability (\ref{eq:2}), being the
square of the amplitude, decays with $\exp(-4 m d)$.}

The important difference to quantum mechanical tunneling is that it is not a
wave function of an on-shell particle that penetrates the
barrier. Instead, the existence of and the interaction with off-shell
intermediate states are necessary to give rise to this new tunneling
phenomenon. This is somewhat similar to the tunneling picture of Schwinger
pair production \cite{Affleck:1981bma,Kim:2000un,Gies:2005bz}: here, the
production of charged particles is facilitated by an external electric field
that assists fluctuations to tunnel out of the vacuum through the spectral gap
to on-shell asymptotic states. Of course, an important difference remains, as
there is a clear distinction in our case between the intermediate fluctuation
states and the asymptotic photons.

Let us now consider the limit of the wall thickness being small compared to
the Compton wavelength of the minicharged particle, $md\ll 1$. Here, the
auxiliary function $f_<$ diverges logarithmically. As this limit probes the
vacuum polarization at larger and larger momentum scales, this logarithmic
behavior corresponds to the logarithmic running of the gauge coupling above
the mass threshold:
\begin{eqnarray}
f_<(md\ll 1) &\simeq& {\ln\left(\frac{1}{2md}\right)}, \nonumber\\
P_{\gamma\to\gamma}(\omega\ll 2m,md\ll 1)
&\simeq &\frac{\alphat^2}{36\pi^2}\, \ln^2(2md)
{=\frac{\alpha^2\charge^4}{36\pi^2}\, \ln^2(2md)}.
\end{eqnarray}
For instance, for a wall thickness in the millimeter range, this limit applies
to minicharged particles with a mass below the meV scale, where the driving
photon frequency $\omega$ is chosen even much below the meV scale.

\subsection{Large-frequency limit $\omega\gg 2 m$}

For larger frequencies $\omega> 2m$, both auxiliary functions $f_>$ and $f_<$
contribute. Here, analytic limits are more difficult to obtain, since
cancelations between the two integrals can occur.

For instance, the limit of a large wall thickness can be obtained from
expanding $f_>$ in Eq.~(\ref{eq:3}) with respect to the upper bound;
contributions from the lower bound are canceled by corresponding contributions
from $f_<$, yielding for the probability
\begin{equation}
P_{\gamma\to\gamma}=\frac{\alpha^2\charge^4\omega^3}{512\pi m^3(dm)^3}
\quad{\rm for}\quad\frac{2m}{\omega}\, (m d)\gg 1.
\end{equation}
In the case of small masses, we note that the approximation is valid only at
fairly large wall thickness, rendering this limit phenomenologically less
relevant.

For not too large $\omega\gtrsim 2m$, the small-wall-thickness limit is
dominated by $f_<$ which reduces to
\begin{equation}
f_<(\omega\gtrsim 2m, \omega d\ll 1) \simeq \int_0^\infty
\frac{d\kappa}{i+\kappa}\, e^{-\omega d \kappa} = e^{i\omega d} \Gamma(0,i
\omega d) \to \ln \left( \frac{1}{\omega d} \right) - \gamma -i \frac{\pi}{2}
+ \mathcal O(\omega d). \label{eq:4}
\end{equation}
Again, the limit of small wall thickness probes the high-momentum structure of
vacuum polarization, yielding a logarithmic increase of the tunneling
probability,
\begin{equation}
  P_{\gamma\to\gamma}(\omega\gtrsim 2m, \omega d\ll1) \simeq \frac{\alpha^2\charge^4}{36\pi^2}\,
  \ln^2 {\left(\frac{1}{\omega d} \right)}.
\end{equation}
Note that the mass $m$ of the fluctuating particle drops out in this limit.

\begin{figure}[t]
\begin{center}
\includegraphics[angle=0,width=.6\textwidth]{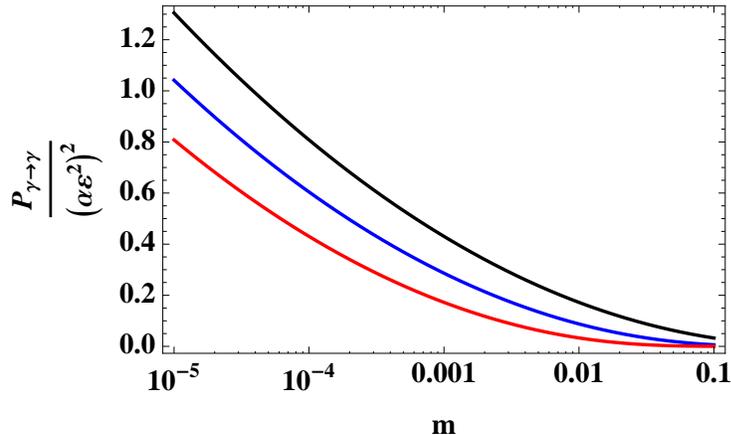}
\end{center}
\vspace{-0cm} \caption{{Transition probability in the large frequency limit for $\omega=1$ and $d=1,10,100$ (from top to bottom; black, blue, red).
For small masses the probability diverges as a logarithm squared as given in Eq.~\eqref{eq:5}.}}
\label{largef}
\end{figure}

Of particular phenomenological interest is the limit $\omega\gg 2m$ for
$\omega d \sim \mathcal O(1)$. This limit is again purely dominated by the
function $f_>$ which develops a logarithmic behavior at the upper bound of the
integral. Numerically, we find
{\begin{equation}
|f_>(\omega\gg 2m, \omega d \sim \mathcal O(1))+f_<(\omega\gg 2m, \omega d \sim \mathcal O(1))| \simeq a\, \ln
\frac{\omega}{2m} -b, \quad a\simeq 2,
\end{equation}
and $b$ is an $\omega d$-dependent offset; e.g., $b\simeq 0,\, 2.3,\, 4.5$ for
$\omega d=1,\,10,\,100$.} To leading order, the tunnel probability is
\begin{equation}
P_{\gamma\to\gamma}(\omega\gg 2m, \omega d \sim \mathcal O(1)) \simeq
\frac{\alpha^2 \charge^4}{36 \pi^2} \, a^2 \ln^2 \frac{\omega}{2m}. \label{eq:5}
\end{equation}
{The probability in this regime is plotted in Fig.~\ref{largef} for the three values $\omega d=1,10,100$.}

\section{Discovery experiments}

Tunneling of the 3rd kind for photons generally leads to a
light-shining-through-a-wall signature.  As discussed in the preceding
section, this signature decreases drastically with increasing thickness of the
wall.  In order to have a chance of observing tunneling of the 3rd kind, a
suitably thin wall is required that provides at the same time for a sufficient
shielding against the ordinary transmission of photons. The higher the photon
energy the bigger the thermal stress on the wall and the greater the
possibility of accidental leakage of photons through the wall. This suggests
either the use of walls which are as perfectly reflecting as possible for a
specifically selected photon frequency or the use of low or zero frequency
photons as, for instance, provided by a constant magnetic field.

\subsection{Optical experiment}\label{sec:optical-experiment}

Let us first consider an experiment of the standard
light-shining-through-a-wall type, where an optical laser is shone against a
wall and a detector for optical photons is placed behind the wall. For our
estimate, we assume that the wall is almost perfectly reflecting for the
photon frequency $\omega$ with zero transmissivity. This may be achieved by
thin-layer optical coating of a thin substrate. For an optical wavelength in
the $\omega\sim \mathcal O(1\mathrm{eV})$ regime, a wall of $\mathcal O(10\dots
100\mu\mathrm{m})$ implying $\omega d \simeq \mathcal O( 10 \dots 100)$ might
be realizable.

This set up is most sensitive for small masses much below the optical
frequency scale. For $\omega\gg 2m$, the outgoing photon rate behind the wall
is given by (cf. Eq.~(\ref{eq:5}) {and Fig.~\ref{largef}})
{\begin{equation}
n_{\mathrm{out}} = n_{\mathrm{in}} P_{\gamma\to\gamma} \simeq 6 \times
10^{-7} \, n_{\mathrm{in}}\, \charge^4 \ln^2 \frac{\omega}{2m},
\end{equation}}
where we have assumed a 100\% detection efficiency.  Even for strong
continuous lasers with {$n_{\mathrm{in}}\sim \mathcal O(10^{20}\dots
  10^{25})/s$}, it is clear that current laboratory {bounds~\cite{Gies:2006ca,Ahlers:2007rd}\footnote{{Astrophysical bounds are even
  stronger, $\epsilon\lesssim 10^{-14}$~\cite{Davidson:2000hf}, but they may be evaded in some models~\cite{Masso:2005ym}.}}} on $\charge$ below
the $\charge\sim 10^{-6}$ range are not immediately accessible, unless the
mass of the minicharged particle is exponentially {small\footnote{{It should
      be noted, however, that for {very small} masses {corresponding
      to Compton wavelengths much larger} than the typical spatial
      dimensions of the experiment in question our approximations, in
      particular the use of plane waves, become unreliable.}}.}

\subsection{Superconducting box experiment}\label{box}

\begin{figure}
\begin{center}
\includegraphics[width=8.6 cm]{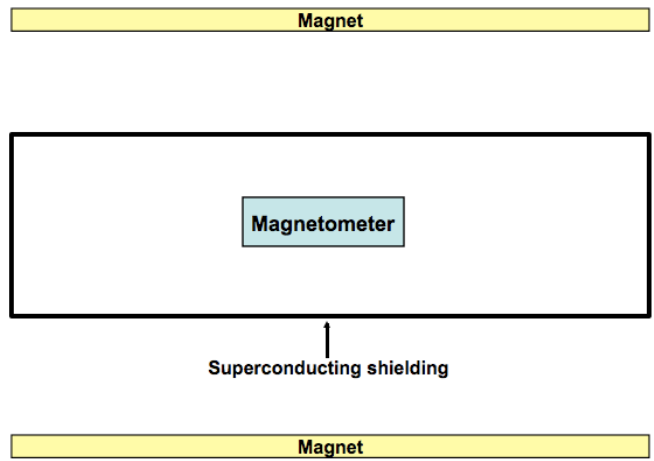}
\end{center}
\vspace{-2ex}
\caption[...]{
Illustration of the principle of {a ``superconducting box'' experiment}. Ordinary magnetic fields are shielded by a Type-I superconductor. However,
the superconductor can be penetrated by the virtual minicharged particle pairs
which generate a non-vanishing magnetic field inside the shielding.
This field can then be measured by a highly sensitive magnetometer.
\label{magnet}}
\end{figure}

{A constant field, i.e. $\omega=0$,} could be realized in the form of a
``superconducting box'' experiment as suggested for the search for
hidden-sector photons in~\cite{Jaeckel:2008sz}.  The basic setup of such an
experiment is depicted in Fig.~\ref{magnet}.  Outside the shielding, we have a
strong magnetic field. Upon entering a Type-I superconductor the ordinary
electromagnetic field is exponentially damped with a length scale given by the
London penetration depth $\lambda_{\rm Lon}$.  Behind the superconducting
shield, the effective current induced by tunneling of the 3rd kind generates a
magnetic field that can be detected by a magnetometer.  Since the magnetometer
measures directly the field (instead of the intensity or power output) the
signal is proportional to the transition amplitude and therefore to the
{coupling} squared, $\charge^2$, instead of being proportional to
$\charge^4$.

{Using} a static magnetic field instead of a wave and replacing the
mirror with the superconductor requires two minor modifications of the
previous calculation.  First, for a static magnetic field, $B_{0}$, impinging on the
superconductor the field on the left hand side of the wall is
constant. Accordingly the $\sin(\omega x)$ in Eq.~\eqref{eq:jright} has to be
replaced by $1$ such that the current {to the right of the wall} now
reads,
\begin{equation}
j_{\rm const}(z>0) = \int_{-\infty}^0 dz'\, \Pi(z-z',\omega)\, {B_{0}}. \label{eq:jright2}
\end{equation}
Second, for a constant field we cannot use the Green's function \eqref{greens}
anymore. {Instead,} it is straightforward to directly solve the
appropriate differential equation,
\begin{equation}
{\partial^{2}_{z}B(z)}=j_{\rm const}(z).
\end{equation}
In the following, we assume that the thickness of the superconducting shielding and the Compton wavelength of the minicharged particle are much
larger than the London penetration depth $d,1/m\gg\lambda_{\rm Lon}$. Then, the relevant part of the induced current is between $d$ and $\infty$.
This implies that $B(d)=-B(\infty)$. The
second required boundary condition is that the field approaches a constant for
$z\to\infty$.\footnote{{Here, we have assumed that the system is
    homogeneous in the $x,y$ direction. Furthermore, the limit of infinite
    extension of this homogeneity has to be taken before the limit
    $z\to\infty$. In a real experiment, the boundary condition can be set at
    large $z$ values which are still smaller than the extent of the field in
    $x,y$ direction.}}  Using this, we find
\begin{equation}
\label{box2}
B(z)-B(d)=\int^{z}_{d}dz^{\prime}\int^{z^{\prime}}_{d}dz^{\prime\prime}j_{\rm const}(z^{\prime\prime})-(z-d)\int^{\infty}_{d}dz^{\prime} j_{\rm const}(z^{\prime}).
\end{equation}

{We can again use the parametrization of the propagator given in
  Eq.~\eqref{eq:Picase}. For a constant $\omega=0$ field only the small
  frequency case contributes.  If we measure the field sufficiently far behind
  the shielding the field strength will be close to the constant asymptotic
  value $B(\infty)$.  Following similar steps as in Sect.~\ref{results}, we
  find for the normal amplitude of the magnetic field behind the wall,
\begin{equation}
\label{boxresult1}
Amp_{\gamma\to\gamma}=\frac{B(\infty)}{B_{0}}=\frac{\alpha\charge^2}{6\pi}g(md),
\end{equation}
where
\begin{equation}
\label{boxresult2}
g(md)=\frac{1}{2}\int^{\infty}_{1}\frac{d\tau}{\tau^{4}}\sqrt{\tau^2-1}(1+2\tau^2)\exp(-2md\tau).
\end{equation}
}

\begin{figure}[t]
\begin{center}
\hspace{-2cm}\includegraphics[angle=0,width=.6\textwidth]{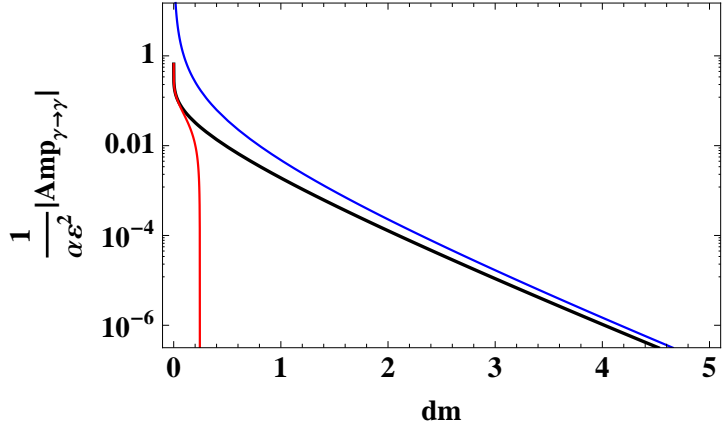}
\end{center}
\vspace{-0cm} \caption{{Amplitude for a constant magnetic field to leak into a volume shielded by a superconductor. The black curve corresponds to a numerical
evaluation of Eq.~\eqref{boxresult1}, {the blue curve gives the approximate result Eq.~\eqref{boxapprox} and the red curve gives the
approximate result Eq.~\eqref{boxsmall}}.}}
\label{boxplot}
\end{figure}

This
amplitude is plotted in Fig.~\ref{boxplot} as a function of the wall thickness
(black line).  For $dm\gg 1$ the transition amplitude can be approximated by,
\begin{equation}
\label{boxapprox}
\left|Amp_{\gamma\to\gamma}\right|
={\left|\frac{B(\infty)}{B_{0}}\right|}={\frac{\alpha\charge^2}{16\sqrt{\pi}}}\frac{\exp(-2dm)}{(dm)^{\frac{3}{2}}}\quad{\rm for}\quad dm\gg 1.
\end{equation}
This is plotted as the blue line in Fig.~\ref{boxplot}.
{For $dm\ll 1$ we find again the appropriate logarithmic divergence
\begin{equation}
\label{boxsmall}
\left|Amp_{\gamma\to\gamma}\right|
={\left|\frac{B(\infty)}{B_{0}}\right|}={\frac{\alpha\charge^2}{6\pi}}\left(\log(dm)+\frac{5}{6}+\gamma\right)\quad{\rm for}\quad dm\ll 1.
\end{equation}
This is shown as the red line in Fig.~\ref{boxplot}.}

Let us now estimate the sensitivity of such an experiment. From
Fig.~\ref{boxplot} we can clearly see that the sensitivity will drop rapidly
if $dm\gg 1$. On the other hand for $dm\lesssim 0.02$ we find
$\left|Amp_{\gamma\to\gamma}\right|>0.1\,\alpha\charge^2$.  Magnetic field
strengths $\Bmag_{0}$ of the order $(1-5)\,{\rm T}$ can be reached in the
laboratory.  However, we have to stay below the critical field strength of the
superconductor in order to prevent penetration of the superconductor by the
magnetic field.  In most materials, this ranges between $0.01\,{\rm T}$ and
$0.2\,{\rm T}$~\cite{Rohlf} although fields as high as $1$ T can be shielded
in certain cases (cf., e.g. \cite{Cavallin}).  Modern magnetometers
\cite{gravityb,robbes,Allred} can detect magnetic fields as low as $\Bmag_{\rm
  detect}=5\times 10^{-18}\, {\rm T}$; {therefore, using $\Bmag_{\rm
    detect}=1\times 10^{-13}$~T} seems relatively conservative.  Accordingly we
can expect a sensitivity in the $\charge\sim 10^{-4}$ to $2\times 10^{-7}$
range\footnote{It should be noted that the shielding of a ${\cal O}(0.1\,{\rm
    T})$ magnetic field down to ${\cal O}(10^{-18}\,{\rm T})$ is certainly an
  experimental challenge. However, shielding on this level has been
  achieved~\cite{gravityb}.}. {The latter is in the ball park of the current best laboratory bounds~\cite{Ahlers:2007rd,Gies:2006ca}.}

Finally, let us find out which mass scales for the minicharged particles we
can probe in the experiment.  Above we have already argued that we can only
achieve good sensitivity as long as $dm\lesssim 0.02$. On the other side we
must have $d\gg\lambda_{\rm Lon}$ in order to suppress the ordinary leakage of
magnetic fields through the superconducting shielding.  Typical London
penetration depths $\lambda_{\rm Lon}=1/M_{\rm Lon}$ are of the order of
$(20-100)\,{\rm nm}$ (cf,. e.g., \cite{Kittel}).  To avoid fields leaking
directly through the shielding (without having to convert into hidden fields)
at the $10^{-20}$ level we need $d\gtrsim 50\, \lambda_{\rm Lon}\sim
(1-5)\,\mu {\rm m}$.  The requirement $m\lesssim 0.02/d $ then allows, in
principle, to search for masses up to $(0.8-4)\,{\rm meV}$.  With {a
  shielding }thicker than the minimal required {size} the experiment will
be sensitive only to smaller masses.

\section{Conclusions}\label{conclusions}
In this note we have discussed a new type of tunneling process of particles
through a barrier.  Whereas ordinary quantum mechanical tunneling allows the
particle to pass through a barrier of finite width and height, field theory
with different particle species can allow particles to circumnavigate a
barrier by converting into a {\emph{real}} particle of a different species
that does not interact with the barrier (cf. Fig.~\ref{classical}).  As argued
in this note, quantum field theory allows to circumnavigate the barrier by
conversion into {\emph{virtual}} particles that do not interact with the wall
as depicted in Fig.~\ref{tunneling}.  As an explicit example of this process
we have calculated the tunneling probability via this ``tunneling of the 3rd
kind'' for the case of photons coupled to minicharged particles.

From a formal perspective, this quantum field theoretic tunneling becomes
reminiscent to quantum mechanical tunneling in the limit where a semiclassical
approximation can be applied to the fluctuating propagators. Here, the
tunneling probability follows a characteristic exponential behavior with an
exponent that increases linearly with the wall thickness. By contrast, for
high frequencies, our tunneling phenomenon has no quantum mechanical analogue
anymore. Contrary to quantum mechanics where any finite barrier is eventually
overcome in a classical sense for increasing energy, our (idealized) wall
remains a potential barrier for all frequencies\footnote{Of course, any real
matter becomes translucent beyond a frequency scale typically set by the
plasma frequency.}. In particular, in the limit of small wall thickness, we
observe a logarithmic increase of the tunneling probability. This dependence
is characteristic for a quantum field theory phenomenon, as it probes the
structure of vacuum polarization at high fluctuation momenta.

Experimentally, tunneling of the 3rd kind could be observed as a
``light-shining-through-a-wall'' signature. In contrast to the process in
classical field theory where real particles traverse the wall, tunneling of
the 3rd kind would lead to a strong dependence on the wall thickness.

\section*{Acknowledgements}
The authors would like to thank Jens Braun for interesting discussions.
HG acknowledges support by the DFG under contract Gi 328/5-1
(Heisenberg program) and by the DFG-SFB TR18.

\begin{appendix}
  \section{Photon transition amplitude using the Feynman-parameter
    representation of the polarization tensor}
As a check of the derivation of the transition amplitude given in the main
text we can use a different representation of the polarization tensor as
given, e.g., in \cite{Peskin:1995ev},
\begin{equation}
\label{poltens}
\Pi(p^2)=-p^2\frac{2\alpha\charge^2}{\pi}\int^{1}_{0}dx\,x(1-x)\log\left(\frac{m^2}{m^2-x(1-x)p^2}\right),
\end{equation}
where the $\charge^2$ accounts for the small charge $\charge$ of the minicharged particles.

We now have to perform the partial Fourier transform of this expression
according to Eq.~\eqref{eq:FTPi} and insert it into
Eq.~\eqref{transition}. The details of this calculation depend on wether we
are at small frequency regime $\omega\ll 2m$ or at large frequencies
$\omega\gg 2m$. We will now study these two cases separately.

\subsection{Small-frequency limit $\omega\ll 2 m$}
The first step in calculating the transition probability is to obtain the
partial Fourier transform \eqref{eq:FTPi} of the polarization tensor given in
\eqref{poltens} for $\mathbf{p}_{\perp}=0$,
\begin{equation}
\Pi(z,\omega)=\int^{\infty}_{-\infty} dp_{z}\,\exp(izp_{z})\Pi(\omega^2-p^{2}_{z}).
\end{equation}

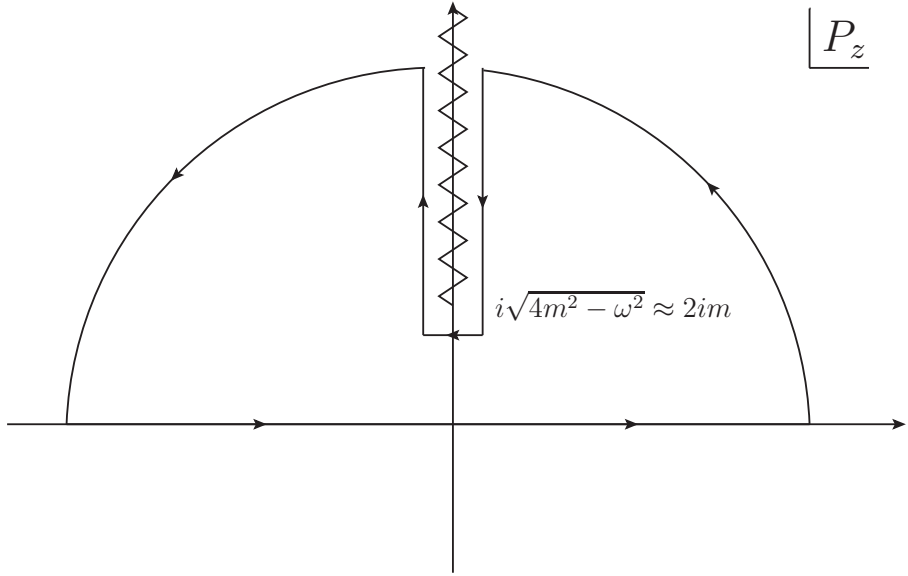
\begin{figure}
\begin{center}
\scalebox{0.7}[0.7]{
\fcolorbox{white}{white}{
  \begin{picture}(500,310) (111,-107)
    \SetWidth{1.0}
    \SetColor{Black}
    \Line[arrow,arrowpos=1,arrowlength=5,arrowwidth=2,arrowinset=0.2](352,-106)(352,198)
    \Line[arrow,arrowpos=1,arrowlength=5,arrowwidth=2,arrowinset=0.2](112,-26)(592,-26)
    \ZigZag(352,38)(352,198){7.5}{8}
    \Line(544,198)(544,166)
    \Line(544,166)(576,166)
    \Arc[arrow,arrowpos=0.5,arrowlength=5,arrowwidth=2,arrowinset=0.2](344,-34)(200.16,2.291,83)
    \Arc[arrow,arrowpos=0.5,arrowlength=5,arrowwidth=2,arrowinset=0.2](344,-34)(200.16,92,177.709)
    \Line[arrow,arrowpos=0.5,arrowlength=5,arrowwidth=2,arrowinset=0.2](368,166)(368,22)
    \Line[arrow,arrowpos=0.5,arrowlength=5,arrowwidth=2,arrowinset=0.2](368,22)(336,22)
    \Line[arrow,arrowpos=0.5,arrowlength=5,arrowwidth=2,arrowinset=0.2](336,22)(336,166)
    \Text(440,38)[c]{\scalebox{1.3}{$i\sqrt{4m^2-\omega^2}\approx 2 i m$}}
    \Text(565,182)[c]{\scalebox{2}{$P_{z}$}}
    \Line[arrow,arrowpos=0.5,arrowlength=5,arrowwidth=2,arrowinset=0.2](144,-26)(352,-26)
    \Line[arrow,arrowpos=0.5,arrowlength=5,arrowwidth=2,arrowinset=0.2](352,-26)(544,-26)
  \end{picture}
}
}
\end{center}
\caption{Integration contour for the small frequency limit $\omega\ll 2m$.} \label{smallmcontour}
\end{figure}

To perform this integration we continue $p_{z}$ into the complex plane $P_{z}$
and integrate along the contour shown in Fig.~\ref{smallmcontour}.  The
relevant parts of this contour are the two bits parallel to the imaginary
axis. Together they contribute
\begin{equation}
\label{largemint}
\Pi(z,\omega)=2\int^{\infty}_{\sqrt{4m^2-\omega^2}}\frac{dP_{z}}{2\pi}\,\exp(-P_{z}z){\rm {Im}}\left(\Pi(\omega^2+P^2+i\epsilon)\right).
\end{equation}
We therefore need the imaginary part of $\Pi(p^2+i\epsilon)$. Using
Eq.~\eqref{poltens} one finds,
\begin{equation}
{\rm Im}\left(\Pi(p^{2}\pm i\epsilon)\right)=\mp\charge^2\frac{\alpha}{3}p^2 \sqrt{1-\frac{4m^2}{p^2}}\left(1+\frac{2m^2}{p^2}\right).
\end{equation}
For $\sqrt{4m^2-\omega^2}\,z\gg1$, we can approximate the integral
\eqref{largemint} by,
\begin{eqnarray}
\Pi(z,\omega)\!\!&=&\!\!-\alpha\charge^2
\frac{m(4m^2-\omega^2)^{\frac{1}{4}}}{\sqrt{2\pi}z^{\frac{3}{2}}}\exp\left(-\sqrt{4m^2-\omega^2}z\right)\quad{\rm for}\quad \sqrt{4m^2-\omega^2}\,z\gg 1
\\\nonumber
\!\!&=&\!\!-\frac{\alpha\charge^2}{\sqrt{\pi}}\frac{m^{\frac{3}{2}}}{z^{\frac{3}{2}}}\exp(-2mz)\quad{\rm for}\quad\omega\to 0.
\end{eqnarray}
The accuracy of this approximation is better than {$15\%$ for
$\sqrt{4m^2-\omega^2}\,z\gtrsim 40$.}  Inserting this expression into
Eq.~\eqref{transition} we find,
\begin{eqnarray}
\label{largemprob}
P_{\gamma\to\gamma}\!\!&=&\!\!\frac{\alpha^2\charge^4}{512\pi}\frac{\sqrt{4m^2-\omega^2}}{d^3m^4}
\exp\left(-2d\sqrt{4m^2-\omega^2}\right)\quad{\rm for}\quad \sqrt{4m^2-\omega^2}\,d\gg 1
\\\nonumber
\!\!&=&\!\!\frac{\alpha^2\charge^4}{256\pi}\frac{\exp(-4dm)}{(dm)^3}\quad{\rm for}\quad \omega\to 0.
\end{eqnarray}
This agrees with the result found in Eq.~(\ref{eq:2}).

\subsection{High-frequency limit $\omega\gg 2 m$}
\begin{figure}
\begin{center}
\scalebox{0.7}[0.7]{
\fcolorbox{white}{white}{
  \begin{picture}(500,310) (111,-107)
    \SetWidth{1.0}
    \SetColor{Black}
    \Line[arrow,arrowpos=1,arrowlength=5,arrowwidth=2,arrowinset=0.2](352,-106)(352,198)
    \Line[arrow,arrowpos=1,arrowlength=5,arrowwidth=2,arrowinset=0.2](112,-26)(592,-26)
    \Line(544,198)(544,166)
    \Line(544,166)(576,166)
    \Arc[arrow,arrowpos=0.5,arrowlength=5,arrowwidth=2,arrowinset=0.2](344,-34)(200.16,2.291,83)
    \Arc[arrow,arrowpos=0.5,arrowlength=5,arrowwidth=2,arrowinset=0.2](344,-34)(200.16,92,177.709)
    \Text(450,10)[c]{\scalebox{1.3}{$\sqrt{\omega^2-4m^2}\approx\omega$}}
    \Text(260,10)[c]{\scalebox{1.3}{$-\sqrt{\omega^2-4m^2}\approx-\omega$}}
    \Text(565,182)[c]{\scalebox{2}{{$P_{z}$}}}
    \ZigZag(352,-26)(352,182){7.5}{10}
    \ZigZag(352,-26)(448,-26){7.5}{5}
    \ZigZag(352,-26)(256,-26){7.5}{5}
    \SetColor{Red}
    \Line[arrow,arrowpos=0.5,arrowlength=5,arrowwidth=2,arrowinset=0.2](144,-26)(240,-26)
    \Line[arrow,arrowpos=0.5,arrowlength=5,arrowwidth=2,arrowinset=0.2](240,-26)(240,-10)
    \Line[arrow,arrowpos=0.5,arrowlength=5,arrowwidth=2,arrowinset=0.2](240,-10)(336,-10)
    \Line[arrow,arrowpos=0.5,arrowlength=5,arrowwidth=2,arrowinset=0.2](336,-10)(368,-42)
    \Line[arrow,arrowpos=0.5,arrowlength=5,arrowwidth=2,arrowinset=0.2](368,-42)(464,-42)
    \Line[arrow,arrowpos=0.5,arrowlength=5,arrowwidth=2,arrowinset=0.2](464,-42)(464,-26)
    \Line[arrow,arrowpos=0.5,arrowlength=5,arrowwidth=2,arrowinset=0.2](464,-26)(544,-26)
    \SetColor{Black}
    \Line[arrow,arrowpos=0.5,arrowlength=5,arrowwidth=2,arrowinset=0.2](336,-10)(336,166)
    \Line[arrow,arrowpos=0.5,arrowlength=5,arrowwidth=2,arrowinset=0.2](368,166)(368,-10)
    \Line[arrow,arrowpos=0.5,arrowlength=5,arrowwidth=2,arrowinset=0.2](368,-10)(464,-10)
    \Line[arrow,arrowpos=0.5,arrowlength=5,arrowwidth=2,arrowinset=0.2](464,-10)(464,-26)
    \SetColor{Blue}
    \Line[arrow,arrowpos=0.5,arrowlength=5,arrowwidth=2,arrowinset=0.2](464,-2)(368,-2)
    \Line[arrow,arrowpos=0.5,arrowlength=5,arrowwidth=2,arrowinset=0.2](369,-48)(465,-48)
  \end{picture}
}
}
\end{center}
\caption{Integration contour for the large frequency limit $\omega\gg 2m$. The
  $i\epsilon$ prescription requires us to integrate along the red
  path. However, to avoid enclosing the cuts we have to close the contour
  along the black curves. We then have to add the two blue parts to recover
  the desired integral.} \label{largecontour}
\end{figure}
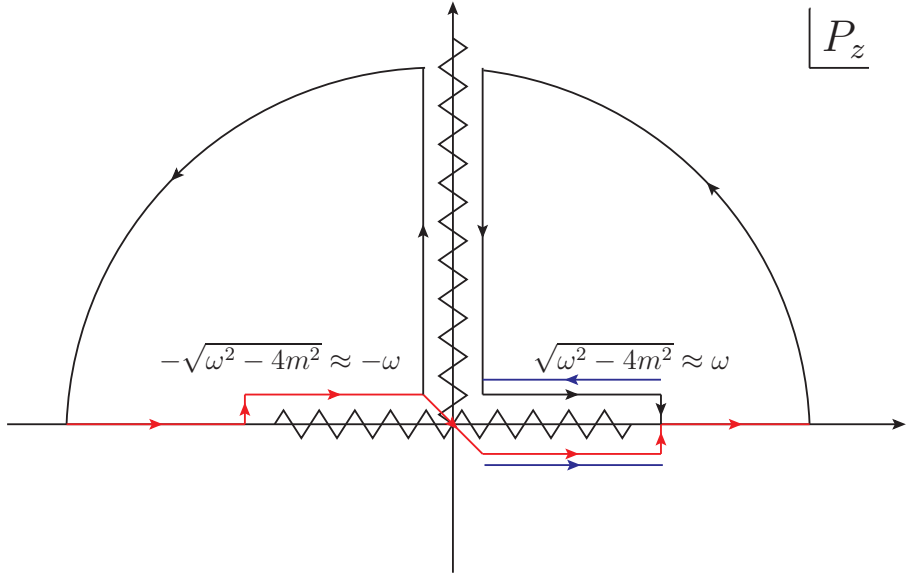

The strategy for the high frequency limit is essentially the same as at low
frequencies.  However, the cut in the complex plane is somewhat more
complicated as can be seen from Fig.~\ref{largecontour}.  The essential
difference to the low frequency case is that the branch cut extends to the
real axis. Therefore, one has to properly take the poles of the propagators
into account.  The usual $i\epsilon$ prescription for the propagators
prescribe that the integration path is along the red line in
Fig.~\ref{largecontour}.  As can be seen from the figure, we have to close the
integration contour along the black paths in order to avoid enclosing the cut
inside the contour.  In order to arrive at the red path, we then have to add
the (finite) blue paths.

The essential contributions to the integral for the Fourier transform are then
again the parts parallel to the imaginary axis and the blue paths,
\begin{eqnarray}
\Pi(z,\omega)=&&\!\!\!\!\!\!\!2\int^{\infty}_{0}\frac{dP_{z}}{2\pi}{\rm Im}\left(\Pi(\omega^2+P^{2}_{z}+i\epsilon)\right)\exp\left(-P_{z}z\right)
\\\nonumber
&&\!\!\!\!\!\!\!\!\!\!\!\!+2i\int^{\sqrt{\omega^2-4m^2}}_{0}\frac{dp_{z}}{2\pi}{\rm Im}\left(\Pi(\omega^2-p^{2}_{z}+i\epsilon)\right)\exp\left(ip_{z}z\right).
\end{eqnarray}
Extracting again the leading order behavior for large distances we find,
\begin{equation}
\Pi(z,\omega)=i\frac{(1+i)}{2\sqrt{\pi}}\alpha\charge^2\frac{m(\omega^2-4m^2)^{\frac{1}{4}}}{z^{\frac{3}{2}}}\exp\left(i\sqrt{\omega^2-4m^2}\,z\right)
\quad{\rm for}\quad\frac{2m^2}{\omega}z\gg 1.
\end{equation}
We note that for small masses the approximation is valid only at fairly large
distances.  Inserting this into Eq.~\eqref{transition}, we obtain the
transition probability,
\begin{equation}
P_{\gamma\to\gamma}=\frac{\alpha^2\charge^4\omega^3}{512\pi
  m^3(dm)^3}\quad{\rm for}\quad\frac{2m^2}{\omega}d\gg 1.
\end{equation}

\end{appendix}


\begin{thebibliography}{10}


\bibitem{bib:tunneling1}
F. Hund, Z. Phys., {\bf 43}, 805 (1927); G. Gamow, Z.\
  Phys.\ {\bf 51}, 204 (1928).

\bibitem{bib:tunneling2}
G. Wentzel, Z.\ Phys.\ {\bf 38}, 518 (1926); H.A. Kramers, Z.\
Phys.\ {\bf 39}, 828 (1926); L. Brillouin, Comptes Rendus {\bf 183}, 24
(1926).

\bibitem{Sikivie:1983ip}
  P.~Sikivie,
  Phys.\ Rev.\ Lett.\  {\bf 51} (1983) 1415
  [Erratum-ibid.\  {\bf 52} (1984) 695];
  A.~A.~Anselm,
  Yad.\ Fiz.\  {\bf 42} (1985) 1480;
  M.~Gasperini,
  Phys.\ Rev.\ Lett.\  {\bf 59} (1987) 396;
K.~Van~Bibber, N.~R. Dagdeviren, S.~E. Koonin, A.~Kerman, and H.~N. Nelson,
Phys. Rev. Lett. {\bf 59}, 759 (1987).

\bibitem{Ehret:2007cm}
  K.~Ehret {\it et al.},
  arXiv:hep-ex/0702023;
  C.~Robilliard,{\it et al}
  [BMV Collaboration],
  Phys.\ Rev.\ Lett.\  {\bf 99} (2007) 190403;
  A.~S.~Chou {\it et al.}  [GammeV (T-969) Collaboration],
  Phys.\ Rev.\ Lett.\  {\bf 100} (2008) 080402;
  P.~Pugnat {\it et al.}  [OSQAR Collaboration],
  arXiv:0712.3362 [hep-ex].
  M.~Fouche {\it et al.} [BMV Collaboration],
  Phys.\ Rev.\  D {\bf 78} (2008) 032013;
  A.~Afanasev {\it et al.} [LIPSS Collaboration],
  arXiv:0806.2631 [hep-ex].

\bibitem{Okun:1982xi}
L.~B. Okun,
Sov. Phys. JETP {\bf 56}, 502 (1982).

\bibitem{Ahlers:2007rd}
  M.~Ahlers, H.~Gies, J.~Jaeckel, J.~Redondo and A.~Ringwald,
  Phys.\ Rev.\  D {\bf 76} (2007) 115005
  [arXiv:0706.2836 [hep-ph]];
  Phys.\ Rev.\  D {\bf 77} (2008) 095001
  [arXiv:0711.4991 [hep-ph]].



\bibitem{Gies:2000wc}
  H.~Gies and R.~Shaisultanov,
  Phys.\ Lett.\  B {\bf 480} (2000) 129
  [arXiv:hep-ph/0009342].

\bibitem{Holdom:1985ag}
  B.~Holdom,
  Phys.\ Lett.\  B {\bf 166} (1986) 196.

\bibitem{Brummer:2009cs}
  F.~Brummer and J.~Jaeckel,
  arXiv:0902.3615 [hep-ph].

\bibitem{Dienes:1996zr}
  K.~R.~Dienes, C.~F.~Kolda and J.~March-Russell,
  Nucl.\ Phys.\  B {\bf 492} (1997) 104
  [arXiv:hep-ph/9610479];
 A.~Lukas and K.~S.~Stelle,
JHEP {\bf 0001} (2000) 010
[arXiv:hep-th/9911156];
  D.~Lust and S.~Stieberger,
  arXiv:hep-th/0302221;
  S.~A.~Abel and B.~W.~Schofield,
  Nucl.\ Phys.\  B {\bf 685}, 150 (2004)
  [hep-th/0311051];
  S.~Abel and J.~Santiago,
  J.\ Phys.\ G {\bf 30}, R83 (2004)
  [hep-ph/0404237];
  R.~Blumenhagen, G.~Honecker and T.~Weigand,
  JHEP {\bf 0506} (2005) 020
  [arXiv:hep-th/0504232];
  B.~Batell and T.~Gherghetta,
  Phys.\ Rev.\  D {\bf 73} (2006) 045016
  [arXiv:hep-ph/0512356];
  R.~Blumenhagen, S.~Moster and T.~Weigand,
  Nucl.\ Phys.\  B {\bf 751} (2006) 186
  [arXiv:hep-th/0603015];
  S.~A.~Abel, J.~Jaeckel, V.~V.~Khoze and A.~Ringwald,
  Phys.\ Lett.\  B {\bf 666} (2008) 66
  [arXiv:hep-ph/0608248];
  S.~A.~Abel, M.~D.~Goodsell, J.~Jaeckel, V.~V.~Khoze and A.~Ringwald,
  JHEP {\bf 0807} (2008) 124
  [arXiv:0803.1449 [hep-ph]].

\bibitem{Jaeckel:2008sz}
  J.~Jaeckel and J.~Redondo,
  Europhys.\ Lett.\  {\bf 84} (2008) 31002
  [arXiv:0806.1115 [hep-ph]].

\bibitem{Dittrich}
  W.~Dittrich and M.~Reuter,
  Lect.\ Notes Phys.\  {\bf 220} (1985) 1.


\bibitem{Schubert:2001he}
  for a review, see C.~Schubert,
  Phys.\ Rept.\  {\bf 355}, 73 (2001)
  [arXiv:hep-th/0101036].

\bibitem{Affleck:1981bma}
  I.~K.~Affleck, O.~Alvarez and N.~S.~Manton,
  Nucl.\ Phys.\  B {\bf 197}, 509 (1982).

\bibitem{Kim:2000un}
  S.~P.~Kim and D.~N.~Page,
  Phys.\ Rev.\  D {\bf 65}, 105002 (2002)
  [arXiv:hep-th/0005078].

\bibitem{Gies:2005bz}
  H.~Gies and K.~Klingmuller,
  Phys.\ Rev.\  D {\bf 72}, 065001 (2005)
  [arXiv:hep-ph/0505099]; 
  G.~V.~Dunne and C.~Schubert,
  Phys.\ Rev.\  D {\bf 72}, 105004 (2005)
  [arXiv:hep-th/0507174]; 
%
  G.~V.~Dunne, Q.~h.~Wang, H.~Gies and C.~Schubert,
  Phys.\ Rev.\  D {\bf 73}, 065028 (2006)
  [arXiv:hep-th/0602176]; 
  R.~Schutzhold, H.~Gies and G.~Dunne,
  Phys.\ Rev.\ Lett.\  {\bf 101}, 130404 (2008)
  [arXiv:0807.0754 [hep-th]].

\bibitem{Gies:2006ca}
  H.~Gies, J.~Jaeckel and A.~Ringwald,
  Phys.\ Rev.\ Lett.\  {\bf 97} (2006) 140402
  [arXiv:hep-ph/0607118];
  A.~Badertscher {\it et al.},
  Phys.\ Rev.\  D {\bf 75} (2007) 032004;
  M.~Ahlers, H.~Gies, J.~Jaeckel and A.~Ringwald,
  Phys.\ Rev.\  D {\bf 75} (2007) 035011
  [arXiv:hep-ph/0612098];
  S.~N.~Gninenko, N.~V.~Krasnikov and A.~Rubbia,
  Phys.\ Rev.\  D {\bf 75} (2007) 075014
  [hep-ph/0612203].

\bibitem{Davidson:2000hf}
  S.~Davidson, S.~Hannestad and G.~Raffelt,
  JHEP {\bf 0005} (2000) 003
  [arXiv:hep-ph/0001179].

\bibitem{Masso:2005ym}
  E.~Masso and J.~Redondo,
  JCAP {\bf 0509} (2005) 015
  [arXiv:hep-ph/0504202];
  J.~Jaeckel, E.~Masso, J.~Redondo, A.~Ringwald and F.~Takahashi,
  hep-ph/0605313;
  Phys.\ Rev.\  D {\bf 75} (2007) 013004
  [arXiv:hep-ph/0610203];
  E.~Masso and J.~Redondo,
  Phys.\ Rev.\ Lett.\  {\bf 97} (2006) 151802
  [arXiv:hep-ph/0606163];
  R.~N.~Mohapatra and S.~Nasri,
  Phys.\ Rev.\ Lett.\  {\bf 98} (2007) 050402
  [arXiv:hep-ph/0610068];
  P.~Brax, C.~van de Bruck and A.~C.~Davis,
  Phys.\ Rev.\ Lett.\  {\bf 99} (2007) 121103
  [arXiv:hep-ph/0703243].


\bibitem{Rohlf}
J.~W.~Rohlf, ``Modern physics from $\alpha$ to $Z_{0}$'', John Wiley \& Sons, Inc, 1994.

\bibitem{Cavallin}
T.~Cavallin, R.~Quarantiello, A.~Matrone and G.~Giunchi,
J. Phys.: Conf. Ser. 43 (2006) 1015-1018.

\bibitem{gravityb}
J.~C.~Mester, J.~M.~Lockhart, B.~Muhlfelder, D.~O.~Murray and M.~A.~Taber,
Advances in Space Research, 25 (2000) 1185. See also http://einstein.stanford.edu/TECH/technology1.html.

\bibitem{robbes}
D.~Robbes, Sens. Actuators A: Phys. {\bf 129} (2006) 86.

\bibitem{Allred}
  J.~C.~Allred, R.~N.~Lyman, T.~W.~Kornack and M.~V.~Roamlis,
  Phys.\ Rev.\ Lett.\  {\bf 89} (2002) 130801.

\bibitem{Kittel}
Ch.~Kittel, ``Introduction to solid state physics'', John Wiley \& Sons, Inc, 2005.


\bibitem{Peskin:1995ev}
  M.~E.~Peskin and D.~V.~Schroeder,
  ``An Introduction To Quantum Field Theory,''
{\it  Reading, USA: Addison-Wesley (1995) 842 p}



\end{thebibliography}
\end{document}